# Modelling self-similar parabolic pulses in optical fibres with a neural network

**Sonia Boscolo [1], John M. Dudley [2] and Christophe Finot [3,\*]**

[1] *Aston Institute of Photonic Technologies, School of Engineering and Applied Science, Aston University, Birmingham B4 7ET, United Kingdom*

[2] *Institut FEMTO-ST, Université Bourgogne Franche-Comté CNRS UMR 6174, Besançon, 25000, France.*

[3] *Laboratoire Interdisciplinaire Carnot de Bourgogne, UMR 6303 CNRS-Université de Bourgogne-Franche-Comté, 9 avenue Alain Savary, BP 47870, 21078 Dijon Cedex, France*

[\*] *Corresponding author:*
E-mail address: *christophe.finot@u-bourgogne.fr*
*Tel.: +33 3 80395926*

**Abstract:** We expand our previous analysis of nonlinear pulse shaping in optical fibres using machine learning [Opt. Laser Technol., 131 (2020) 106439] to the case of pulse propagation in the presence of gain/loss, with a special focus on the generation of self-similar parabolic pulses. We use a supervised feedforward neural network paradigm to solve the direct and inverse problems relating to the pulse shaping, bypassing the need for direct numerical solution of the governing propagation model.

**Keywords:** pulse shaping, neural networks, nonlinear propagation, parabolic pulse, optical fibers.

# I. Introduction

Machine learning is transforming the scientific landscape, with the use of advanced algorithmic tools in data analysis yielding new insights into many areas of fundamental and applied science [1]. Photonics is no exception [2-4], and machine-learning methods have been applied in a variety of ways to optimise and analyse the output of optical fibre systems. For example, from a control and feedback perspective, various machine-learning strategies have been deployed to design and optimise mode-locked fibre lasers [5-9], optimise optical supercontinuum sources [10], analyse the complex nonlinear dynamics occurring during the buildup of supercontinuum [11-13], and control pulse shaping [14].

In parallel with these developments, pulse shaping based on nonlinear propagation effects in optical fibres has developed into a remarkable tool to tailor the spectral and temporal content of light signals [15, 16], leading to the generation of a large variety of optical waveforms such as ultra-short compressed pulses [17], parabolic- [18], triangular- [19, 20] and rectangular- [21] profiled pulses. Very different features of nonlinear pulse evolution can be observed depending on the sign of the fibre group-velocity dispersion (GVD), which results in specific changes of the pulse temporal shape, spectrum and phase profile. Specifically, the temporal and spectral features of pulses propagating in fibres with anomalous GVD are typically governed by soliton dynamics. Conversely, because the nonlinear dynamics of pulses propagating in fibres with normal GVD are generally sensitive to the initial pulse conditions, it is possible to nonlinearly shape conventional laser pulses into various specialised waveforms through control of the initial pulse temporal intensity and/or phase profile [15]. The presence of gain or losses can further modify the nonlinear propagation dynamics. For instance, propagation in a normally dispersive fibre with distributed gain leads to the attraction of the temporal and spectral properties of any initial pulse towards a parabolic intensity profile which is then maintained with further propagation. Parabolic self-similar pulses in fibre amplifiers are, along with solitons in passive fibres, the most well-known classes of nonlinear attractors for pulse propagation in optical fibre [22, 23]. The unique properties of parabolic similaritons have stimulated numerous applications ranging from high-power ultrashort pulse generation to optical nonlinear processing of telecommunication signals. Yet, due to the typically large number of degrees of freedom involved, optimising nonlinear pulse shaping for application purposes may require extensive numerical simulations based on the integration of the nonlinear Schrödinger equation (NLSE) or its extensions. This is computationally demanding and potentially creates a severe bottleneck in using numerical techniques to design and optimise experiments in real time.

In our previous work [24], we presented a solution to this problem using a supervised machine-learning model based on a feedforward neural network (NN) to solve both the direct and inverse problems relating to pulse shaping in a passive, lossless fibre, bypassing the need for numerical solution of the governing propagation model. Specifically, we showed how the network accurately predicted the temporal and spectral intensity profiles of the pulses that form upon nonlinear propagation in fibres with both normal and anomalous dispersion. Further, we demonstrated the ability of the NN to determine the nonlinear propagation properties from the pulses observed at the fibre output, and to classify the output pulses according to the initial pulse shape. In this paper, we extend the use of our model-free method and show that a feedforward NN can excellently predict the behaviour of nonlinear pulse shaping in the presence of distributed gain or loss. The network is able to infer the key characteristics of parabolic pulses with remarkably high accuracy, and to successfully handle variations of the pulse parameters over more than two orders of magnitude.

## II. Governing propagation model and neural network

### Situation under study, model and analytical results

Optical pulse propagation in an optical fibre with distributed gain/loss can be well described by a modified NLSE including the effects of linear GVD, nonlinear self-phase modulation (SPM) and linear gain/loss [25]. Using the dimensionless variables $u = \psi / \sqrt{P_0}$, $\xi = z/L_D$ and $\tau = t/T_0$, this equation is written as

$$i\frac{\partial u}{\partial \xi} - i\frac{\delta}{2}u - \frac{1}{2}\frac{\partial^2 u}{\partial \tau^2} + N^2 |u|^2 u = 0 \qquad (1)$$

Here, $\psi(z,t)$ is the complex electric field envelope in a comoving system of coordinates, $T_0$ and $P_0$ are a characteristic temporal value and the peak power of the initial pulse, respectively, $L_D = T_0^2/|\beta_2|$, $L_{NL} = 1/(\gamma P_0)$ and $N = \sqrt{L_D / L_{NL}}$ are the dispersion length, the nonlinear length and the 'soliton-order' number (or power parameter), respectively, $\delta = \alpha L_D$, and $\beta_2$, $\gamma$ and $\alpha$ are the respective GVD, Kerr nonlinearity and gain/loss coefficients of the fibre, where $\alpha > 0$ ($\alpha < 0$) describes distributed gain (loss). We consider here propagation at normal dispersion (characterized by $\beta_2 >0$), and we neglect higher-order linear and nonlinear effects as the leading-order behavior is well approximated by Eq. (1) for picosecond pulses. Moreover, neglecting higher-order gain effects is well-suited to describe experiments that use broadband rare-earth fibre amplifiers. We note that the dimensionality reduction entailed by Eq. (1) i.e., the mapping of the nonlinear shaping problem from the six-dimensional space of physical parameters ($T_0$, $P_0$, $\beta_2$, $\gamma$, $\alpha$, $L$) where $L$ is the fibre length to the three-dimensional space of ($\xi$, $N$, $\delta$), relaxes the complexity of the analysis, and for a specific selected set of normalized parameters there are many groups of physical parameters suitable the defining equations of $\xi$, $N$ and $\delta$. Different initial pulse shapes are studied: a Gaussian pulse $u_0(\tau) = \exp(-\tau^2/2)$, a hyperbolic secant pulse $u_0(\tau) = \mathrm{sech}(\tau)$, a parabolic pulse $u_0(\tau) = \sqrt{1-\tau^2}\ \theta(1-|\tau|)$, and a second-order super-Gaussian pulse $u_0(\tau) = \exp(-\tau^4/2)$. Here, $\theta(x)$ is the Heaviside function.

It is useful to review here the main features of the pulse evolution in a normally dispersive fibre as described by Eq. (1) in the nonlinearity-dominant regime of propagation, i.e. when $N \gg 1$. In the absence of gain, during the initial stage of propagation, the combined action of SPM and normal GVD makes a standard laser pulse (such as a Gaussian or hyperbolic secant waveform) acquire a frequency chirp with a linear variation over the pulse center, which results in the broadening of the pulse and reshaping into a convex-up parabola. However, the parabolic shape formed is not maintained with propagation, but the pulse evolves toward a near-trapezoidal form with a linear frequency variation over most of the pulse and ultimately, when the shifted light overruns the pulse tails, the wave breaks and develops oscillations on its edges [26]. After the distance at which wave breaking occurs, and with the accumulation of a parabolic spectral phase induced by dispersion, the temporal and spectral content of the pulse become increasingly close to each other and the temporal intensity profile does not evolve anymore. This long-term far-field evolution corresponds to the formation of a spectronic nonlinear structure in the fibre [27].

The situation is different for an intense pulse with an initial parabolic profile, which keeps its shape and acquires a linear chirp upon propagation in the fibre, thereby avoiding the degrading effect of wave breaking [28]. Moreover, in contrast to passive propagation where the parabolic waveform that develops after initial propagation represents a transient state of the pulse evolution over a distance depending on the initial conditions [29], the presence of gain in the fibre results in any arbitrary pulse asymptotically reshaping into a parabolic intensity profile, and this pulse form is maintained with further propagation. The dynamic evolution of the intensity of any pulse in the limit $\xi \rightarrow \infty$ can be in fact accurately described by [22]

$$I_{PAS} = \left(1 - \tau^2 / \tau_P^2\right) \theta\left(\tau_P - |\tau|\right) \tag{2}$$

where the dynamic equation for the characteristic width $\tau_P$ has the explicit solution

$$\tau_P^3 = 3^3 \frac{\sqrt{\pi}}{2} \left(\frac{N}{\delta}\right)^2 \exp(\delta \xi). \tag{3}$$

## Numerical simulation and neural networks employed

The data from numerical simulations of the NLSE (1) is used to train a NN and validate its predictions. Equation (1) is solved with a standard split-step Fourier propagation algorithm [25] using a uniform grid of $2^{14}$ points on a time window of length 280 $T_0$. The latter is chosen wide enough to accommodate the large temporal broadening experienced by the pulses at normal GVD. Our simulations do not include any source of noise such as, e.g., quantum noise. Propagation in the fibre is studied up to a normalized distance of $\xi = 8$, for input powers $N$ up to 4.5 and $\delta$ ranging from –0.36 to 0.8 (i.e. covering both loss and gain). Our main interest here is in temporal and spectral intensity profiles ($I =$ and $S$, respectively) that can be directly recorded in experiments, rather than in the complex envelope of the electric field. As we decide to normalize both $I$ and $S$ by their peak value, information regarding the pulse energy is lost. The simulation pulse profiles are then anamorphically sampled along the temporal and spectral dimensions so as to capture both the details of the short pulses observed at relatively short distances in the fibre and the much longer pulses that are formed at larger distances. Moreover, exploiting the symmetry of the problem, we restrict our sampling to the positive times and frequencies only. Hence, we consider $A = 90$ points on the interval 0 to 150 $T_0$ to represent the temporal intensity profiles of the pulses, and $B = 60$ points on the interval 0 to $5/T_0$ for the spectral intensities. Note that the total number of data points $A+B$ is significantly larger than that used in our previous study [24]. The sampled data with initial conditions taken randomly are used to train a NN and validate its predictions.

We employ a feedforward NN structure relying on the Bayesian regularization back propagation algorithm and including three hidden layers with eighteen, fourteen and ten neurons, respectively, as shown in Fig. 1. This NN is programmed in Matlab using the neural network toolbox. The NLSE numerical simulations are performed on a computer cluster (one and half days is approximately required to produce $25\times10^3$ data samples when 12 processor cores are used). The training and use of the NN are achieved on a standard personal computer (Intel Xeon processor, 6 processor cores, 3.6 GHz, 32 GB RAM).

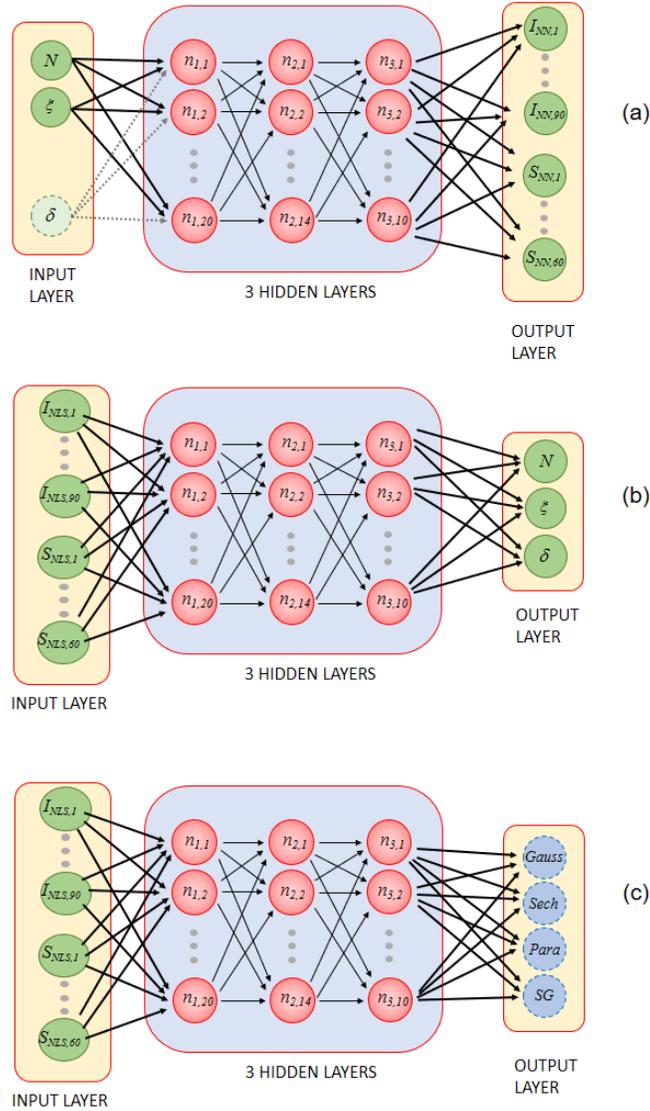

Figure 1: NN model used to: (a) predict the output pulse intensity profiles from the fibre, (b) retrieve the nonlinear propagation properties and (c) identify the input pulse waveform.

# III. Model-free modelling of nonlinear pulse shaping

## Prediction of the output pulse properties

The direct-problem NN learns the NLSE model from 25000 numerical simulations of the propagation of a transform-limited Gaussian pulse with randomly chosen combinations of $\xi$, $N$ and $\delta$. After training, the network is tested on $10^5$ new simulations from random initial conditions. Figure 2 shows the temporal and spectral profiles of the pulse obtained from the network for $\xi = 8$ and $N = 4$ when the fibre exhibits loss ($\delta = -0.23$) or gain ($\delta = 0.69$). The predictions from the NN algorithm show excellent agreement with the results of the NLSE model over the entire temporal pulse shape, and the associated spectrum. In particular, the network is able to reproduce the large temporal and spectral broadening experienced by the pulse upon propagation. For lossy propagation, the output temporal and spectral profiles feature a triangular-like shape [20]. On the contrary, the output pulse from the gain fibre has a parabolic form with rather compact support in both the temporal and spectral domains, as expected from the attracting nature of amplifier similaritons.

To study the accuracy of the NN predictions, we also train a second NN with simulated temporal profiles now expressed in dB units and recorded over a 250 $T_0$-long time window, which entails an intensity dynamic range of 60 dB. The bottom parts of panels (a) show that this network can excellently reproduce the decay of the wings about the pulse core region over several orders of magnitude. Note that for propagation in the presence of gain, these low-amplitude wings have a linear nature (when plotted on a logarithmic scale), and are associated with the intermediate asymptotic regime of the self-similar propagation [30, 31].

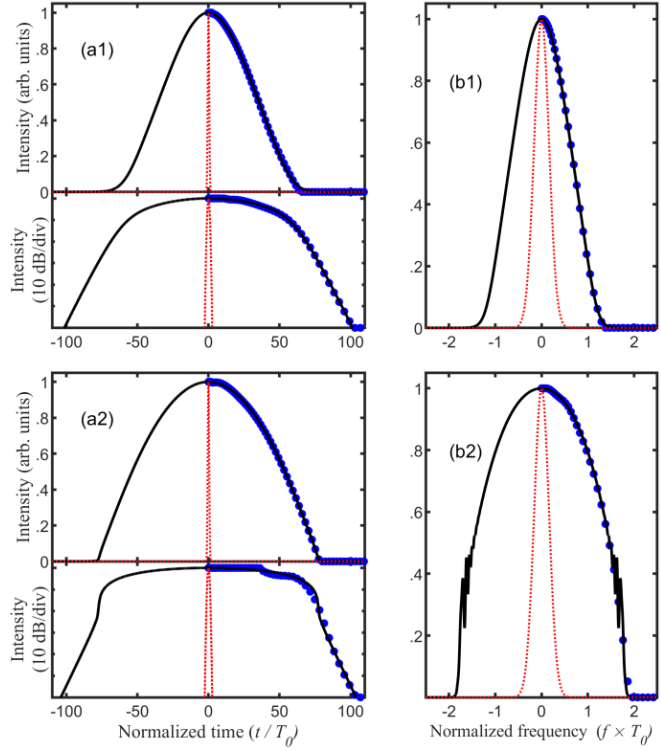

Figure 2: (a) Temporal and (b) spectral intensity profiles of an initial Gaussian pulse after propagation in a lossy/gain normally dispersive fibre with $\xi = 8$, $N = 4$ and $\delta = -0.23$ (panels 1) or $\delta = 0.69$ (panels 2). The predictions from the NN algorithm (blue circles) are compared with the results of NLSE numerical simulations (black curves). Also shown are the input intensity profiles (red dotted curves). The temporal profiles predicted by a second NN are displayed in the bottom parts of panels a using a logarithmic scale.

The prediction error of the NN algorithm is quantified by the parameter of misfit between the output temporal or spectral intensity profile generated by the network, $I_{NN}/S_{NN}$, and the expected profile produced by numerical simulation of the NLSE, $I_{NLS}/S_{NLS}$, given by [24]

$$M_T^2 = \sum_{i=1}^{A}\left(I_{NN,i} - I_{NLS,i}\right)^2 \Big/ \sum_{i=1}^{A} I_{NLS,i}^2$$
$$M_F^2 = \sum_{i=1}^{B}\left(S_{NN,i} - S_{NLS,i}\right)^2 \Big/ \sum_{i=1}^{B} S_{NLS,i}^2$$

(4)

where the expected profiles are interpolated to the same time or frequency points used for sampling the network output. The maps of misfit parameter values in the space of input parameters ($\xi, N, \delta$) are given in Fig. 3(a) and confirm that the nonlinear pulse shaping occurring in the fibre is very closely reproduced by our NN. The distributions of the misfit parameter values [Fig. 3(b)] show that, remarkably, more than 92% of the error realizations are well confined to values below 0.02. These results are significantly better than those reported in [24], where the anomalous dispersion regime of the fibre was more complex to deal with.

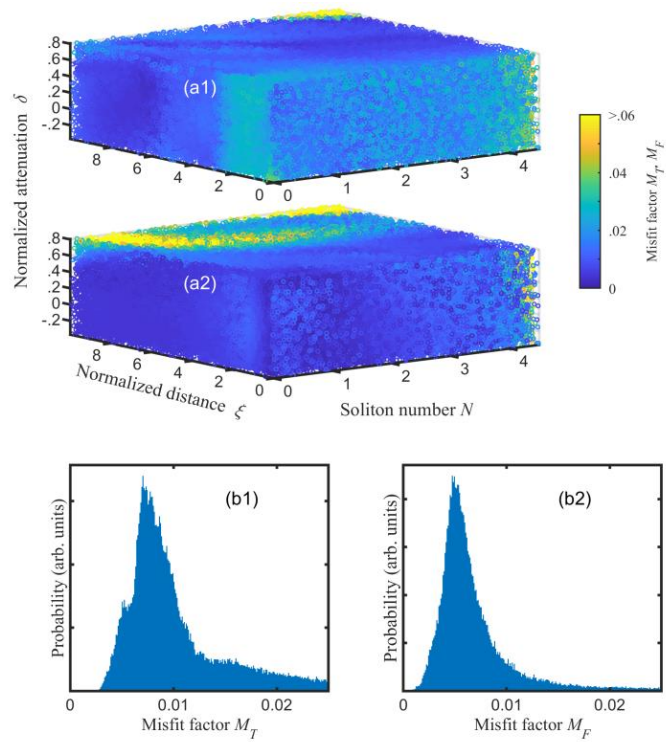

Figure 3: (a) Maps of misfit parameter values between the NN predictions and the NLSE simulation results for the output pulse shape (subplot a1) and optical spectrum (subplot a2) in the three-dimensional space ($N, \xi, \delta$), relating to the propagation of an initial Gaussian pulse in a lossy/gain normally dispersive fibre with randomly chosen combinations of ($N, \xi, \delta$). (b) Distributions of the temporal (subplot b1) and spectral (subplot b2) misfit parameter values.

To further assess the prediction ability of the direct-problem NN, we also compare the temporal and spectral widths (FWHM) of the output pulse as measured from NLSE simulations, which rely on a very fine time and frequency discretization ($2^{14}$ points), with those extracted by the NN from the reduced set of sampling points (only 150 unequally spaced points). The results are summarized in Fig. 4(a), and highlight that even though the pulse temporal duration varies by more than two orders of magnitude across the space of input parameters, the network is able to retrieve this characteristic with a remarkable level of confidence. The spectral width is also predicted accurately. Further to this, we also compute the excess kurtosis [32] (defined as $\mu_4/\sigma^4 - 3$ where $\mu_4$ and $\sigma$ are the fourth-order central moment and standard deviation of the pulse intensity profile, respectively, and 3 is the kurtosis of a Gaussian profile) to characterize the change in the pulse shape resulting from nonlinear propagation in the fibre. It is worth recalling that the interaction between normal dispersion and Kerr nonlinearity causes the reshaping of a propagating pulse into a platykurtic waveform in which the excess kurtosis value is negative. Some deviation of the NN predictions form the NLSE simulation results appears when dealing with the excess kurtosis parameter [Fig. 4(b)] because of the difficulty encountered by the NN in correctly evaluating the fourth moment of the pulse intensity. Yet, comparing the results achieved for propagation lengths longer than 1 (inset plots in Fig. 4(b)) with the full set of data, we can see that the most important discrepancy occurs during the first stage of the nonlinear propagation. We explain this as partly due to a stronger impact of the limited number of sampling points fed to the network on the computation accuracy for the shorter pulses. Finally, we note that the use of a NN that we make in this work is different from and more powerful than that presented in [14], where the NN was trained using the temporal and spectral pulse durations and kurtosis parameter directly.

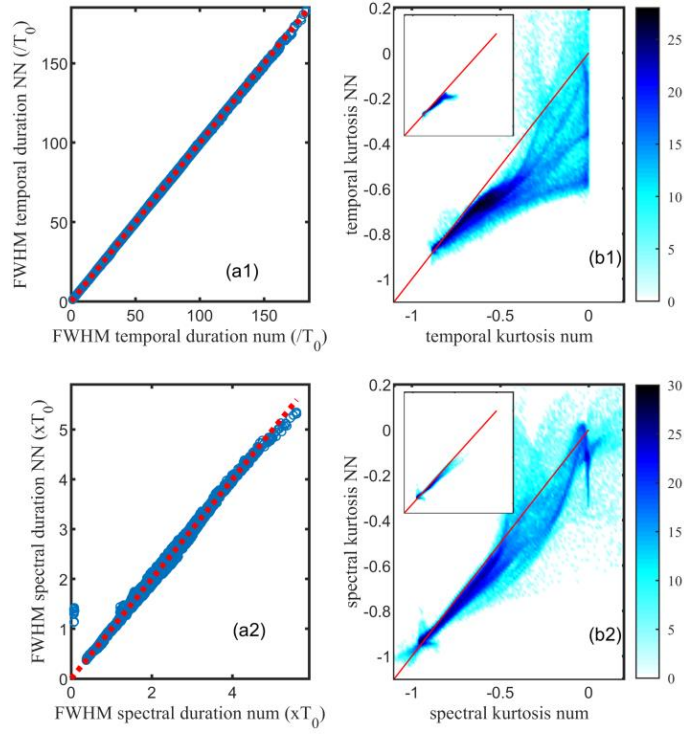

Figure 4: Regressions between the temporal (panels 1) and spectral (panels 2) FWHM duration and excess kurtosis values (panels a and b, respectively) predicted from the NN algorithm and the expected values from the simulated data, relating to the propagation of an initial Gaussian pulse in a lossy/gain normally dispersive fibre with randomly chosen combinations of ($N, \xi, \delta$). Each FWHM duration data point is plotted with a blue circle, while the excess kurtosis data points are colored by density (in dB units). The excess kurtosis results associated with propagation lengths $\xi$ >1 are plotted as insets (with the same axes and color scale as the main graphs).

## Parabolic pulse properties

We now assess the ability of the network to identify the region of the parameter space ($\xi$, $N$, $\delta$) that enable the formation of a parabolic waveform in the fibre. To this end, the trained NN is interrogated with $10^6$ new simulations covering the whole input parameter space and asked to isolate the region that leads to the desired output pulse form by using the excess kurtosis of the calculated intensity profiles as a measure of shape. An excess kurtosis between –0.9 and –0.8 is typical of a parabolic intensity distribution. We note that the NN performs this task in only a few seconds. We can see in Fig. 5(a) that a parabolic pulse is found in two different regions of the parameter space. The region featuring short propagation lengths and high soliton-number values corresponds to the transient state of the nonlinear dynamic pulse evolution toward wave breaking in a passive fibre, and to the early stage of evolution toward the asymptotic solution in a gain fibre. This reshaping stage has been discussed in [26, 33] and used in several experiments to produce parabolic profiles [34, 35]. The region featuring large propagation lengths and $\delta > 0$ underpins the asymptotic parabolic regime in a fibre amplifier. The results relating to the spectral shape of the pulse shown in Fig. 5(b) highlight the different nature of the dynamics enabled by the two regions. Whereas the second region also supports a parabolic pulse form in the spectral domain, this is not the case for the first region. Note that the formation of parabolic spectronic pulses in a passive fibre [27, 36, 37] with a Gaussian pulse initial condition falls beyond the boundary of the parameter space being studied.

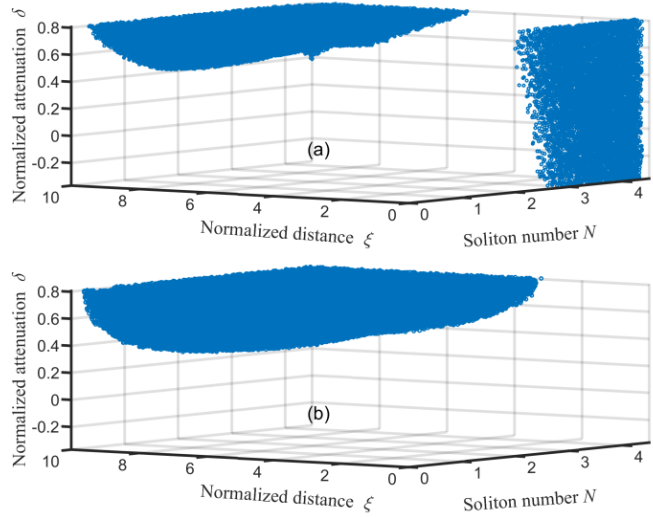

Figure 5: Regions in the three-dimensional space $(N, \xi, \delta)$ that enable the formation of a parabolic intensity profile (characterized by an excess kurtosis value between –0.9 and –0.8) in the (a) temporal and (b) spectral domains, as obtained from the NN model.

Figure 6(a) shows the map of temporal FWHM pulse duration values as computed by the NN from the predicted output pulse profiles over the full search parameter space. We notice again the large variation of this pulse characteristic across the space of input parameters and, expectedly, the fact that the largest pulse durations are enabled by the combination of the highest values of $\xi$, $N$ and $\delta$.

It is also interesting to compare the NN results obtained for positive $\delta$ values with the theoretical predictions from the asymptotic analysis $\tau_{FWHM} = \sqrt{2}\,\tau_P$ with $\tau_P$ given by Eq. (3)]. The relative change of the pulse duration predicted by the NN with respect to the asymptotic values is shown in Fig. 6(b), and confirms that longer propagation length and higher gain enables pulse properties that increasingly approach the asymptotic limits. This difference, however, may still affect inverse design problems. Owing to the three-dimensional representation of the shaping problem being studied, we can deploy a graphical method [14] and, based on the NN predictions, isolate the region of input parameters that supports a given large pulse duration, e.g., $\tau_{FWHM} = 100$. As it is seen in Fig. 6(c), the enabling parameters are not limited to a single combination but are spread over a large surface area. The quite significant deviation of this region from that obtained from Eq. (3) indicates that the asymptotic analysis may not be an efficient tool for solving inverse problems.

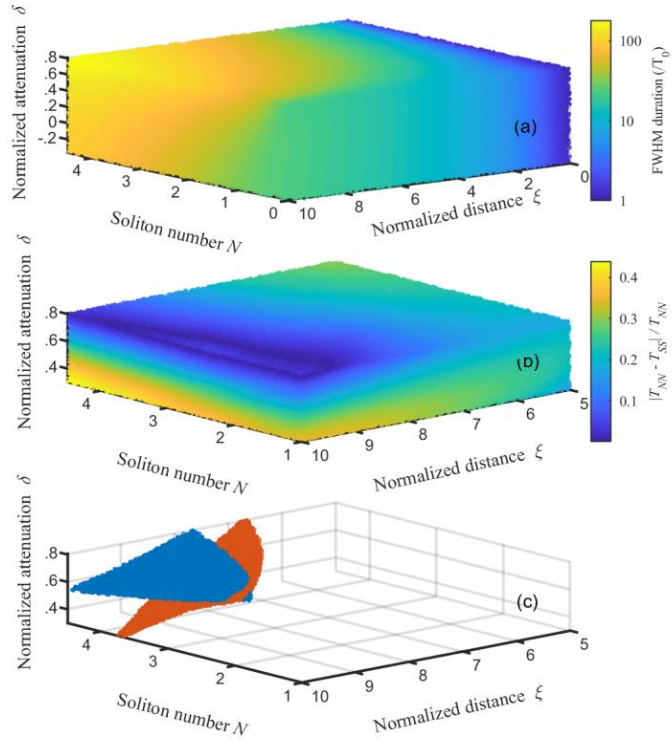

Figure 6: (a) Map of output FWHM pulse duration values obtained from the NN model in the three-dimensional space $(N, \xi, \delta)$. (b) Map of relative change of the pulse duration values predicted by the NN with respect to the values predicted from the asymptotic analysis [Eq. (3)] in the space $(N, \xi, \delta > 0)$. (c) Surfaces in the space $(N, \xi, \delta > 0)$ that enable an output pulse duration of $\tau_{FWHM} =$

100. The results generated from the NN model (blue surface) are compared with the results obtained from the asymptotic analysis (red surface).

Furthermore, we use our NN model to study the evolution of pulses with different initial shapes in a gain fibre. To this end, we train four NNs on ensembles of simulated output pulses from the fibre corresponding to Gaussian, hyperbolic secant, parabolic and super-Gaussian initial pulse shapes, respectively, and random combinations of input parameters. We consider here pulses with the same characteristic width $T_0$=2 ps and energy of 26 pJ inserted into a 5.9 km-long section of fibre ($\xi$=10) with the parameters: $\beta_2$ = 4.89 ps$^2$/km, $\gamma$ = 2.23 /W/km and integrated gain of 34.7 dB ($\delta$ = 0.8). The resulting $N$ values are 4, 3.1, 3.9 and 3.6 for the Gaussian, hyperbolic secant, parabolic and super-Gaussian pulses, respectively. The input pulse duration and the fibre dispersion and nonlinearity parameters correspond to those used in our previous experiments with distributed Raman amplifiers [38]. The initial pulse profiles shown in Fig. 7(a) show a non-negligible difference in terms of pulse duration and peak power. Figure 7(b) illustrates the convergence of the initial pulses towards the theoretically predicted asymptotic state as obtained from the NN model. The output pulse profiles shown in Fig. 7(c) (obtained from NLSE numerical simulations) further confirm that the properties of the asymptotic parabolic solution do not depend on the shape or duration of the initial pulses, but they are solely determined by the initial pulse energy and the fibre parameters.

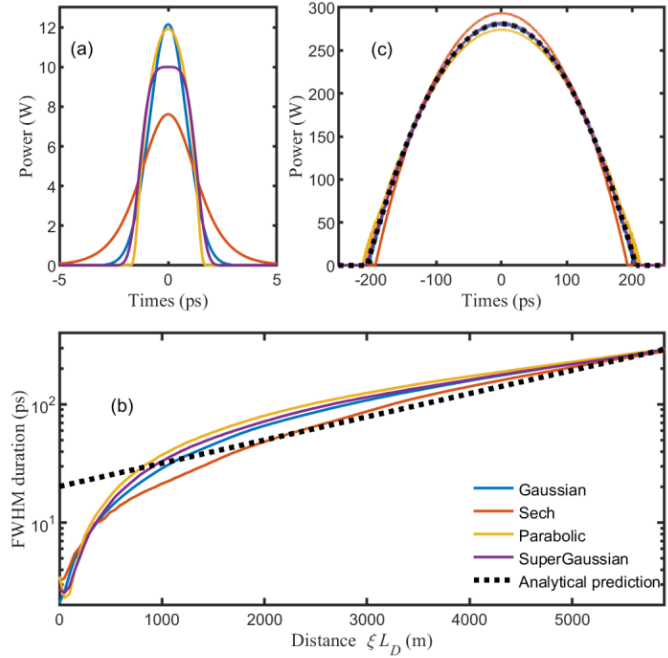

Figure 7: Pulse evolution towards the asymptotic attracting solution in a Raman fibre amplifier for initial Gaussian, hyperbolic secant, parabolic and super-Gaussian pulses (blue, red, yellow and purple solid curves, respectively) with the same energy. (a) Temporal intensity profiles of the input pulses. (b) Evolution of the FWHM temporal duration. The results generated from the NN model are compared with the theoretical predictions [Eq. (3), black dashed curve]. (c) Temporal intensity profiles of the output pulses from the amplifier. The results obtained from NLSE numerical simulations are compared with the theoretical predictions [Eq. (2), black dashed curve].

## Retrieval of the propagation characteristics

For the inverse problem at hand, the trained network is tasked with the retrieval of the propagation parameters $\xi$, $N$ and $\delta$ from a pulse shape and spectrum generated after propagation in the fibre. As we noted in [24], the use of a reverse propagation method [39, 40] to solve this problem would require proper knowledge of the peak power (or energy) of the pulse at the fibre output, whereas we only consider here the pulse shape characteristics. For the NN training phase, we use the same data set as that used for the direct problem. The results of testing the trained NN on $10^5$ randomly chosen new simulated output pulses are shown in Fig. 8. We can see from Fig. 8(a) that the estimation error on $\xi$ ($\Delta\xi$; defined as the difference between the retrieved parameter value and the target value extracted from the NLSE simulation data) is well below 0.1 for all test realizations except those corresponding to high gain values and large propagation lengths. This parameter region corresponds to the propagation stage in which a pulse approaches the asymptotically evolving parabolic solution in the fibre amplifier, where any pulse with a given initial energy would converge to the same intensity profile, thereby leading to the same shape for different $\xi$ values. Outside this region, interestingly, each temporal and spectral shape can be unambiguously associated with a single parameter set ($\xi$, $N$, $\delta$). This is quite remarkable if we consider that several different combinations of parameters may, e.g., lead to the same pulse temporal duration [see Fig. 6(c)]. The regressions between the predicted values of $N$, $\xi$ and $\delta$ from the NN algorithm and the exact target values from the simulated data [Fig. 8(b)] indicate that the gain/attenuation parameter is most sensitive to the incurrence of out of range values. The histograms of the estimation errors shown in Fig. 8(c) confirm the remarkable accuracy of the results obtained with the NN algorithm. Their central parts can be represented by normal distributions of standard deviations $3.8 \cdot 10^{-3}$, $5.8 \cdot 10^{-3}$ and $1 \cdot 10^{-3}$ for $\xi$, $N$ and $\delta$, respectively.

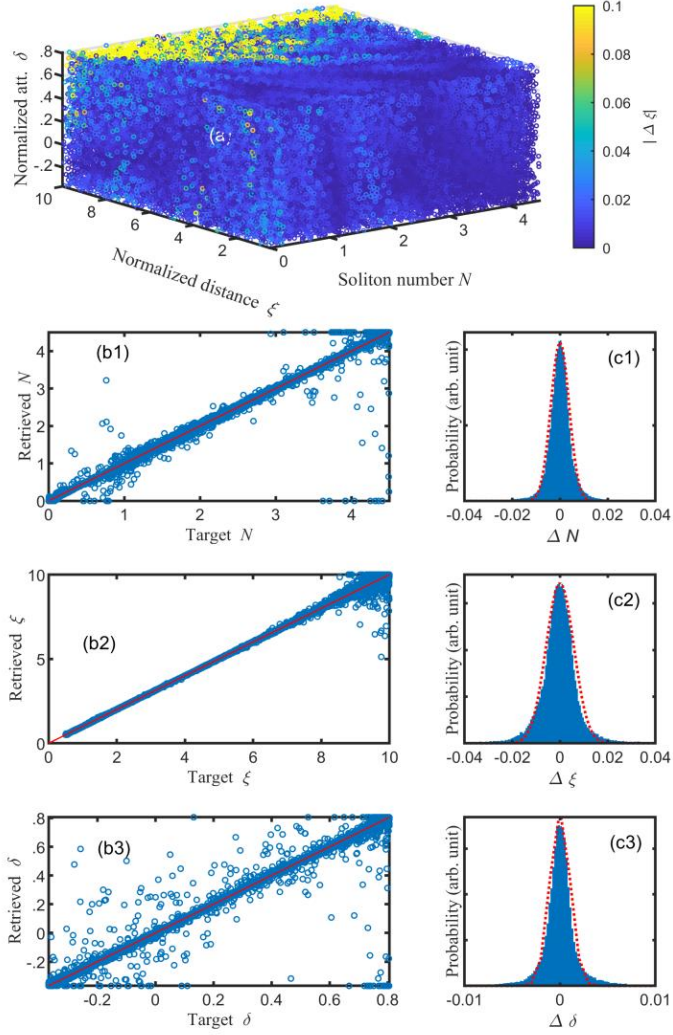

Figure 8: (a) Map of estimation error values on the normalized propagation length $\xi$ in the three-dimensional space ($N$, $\xi$, $\delta$), when the NN is interrogated with randomly chosen simulated output pulses from the fibre. (b) Regressions between the predicted values of $N$, $\xi$ and $\delta$ from the NN algorithm and the exact target values from the simulated data. (c) Distributions of the estimation errors on $N$, $\xi$ and $\delta$. The histograms are compared to normal probability density functions (red dotted curves).

## Identification of the initial pulse shape

For this problem, we train the network on an ensemble of 36000 simulated output pulses from the fiber corresponding to a mix of Gaussian, hyperbolic secant, parabolic and super-Gaussian initial pulse shapes and randomly chosen combinations of input parameters $\xi$, $N$ and $\delta$. Then we ask the trained network to categorize $4 \times 10^5$ new unlabeled simulated output pulses according to the initial waveform. As we can observe from Fig. 9(a), the classification accuracy of the NN algorithm is remarkably high: there are only 859 errors, which represent 0.2% of the total number of input samples. Figure 9(b) indicates that once again the errors occur in the parameter region corresponding to the onset of the asymptotic parabolic regime in a fibre amplifier [see Fig. 5(a) and Fig. 6(b)]. Indeed, as it is also seen in Fig. 7(c), the process of attraction towards the asymptotic parabolic shape tends to conceal the impact of the initial condition, thereby making the waveform recognition more challenging for the network.

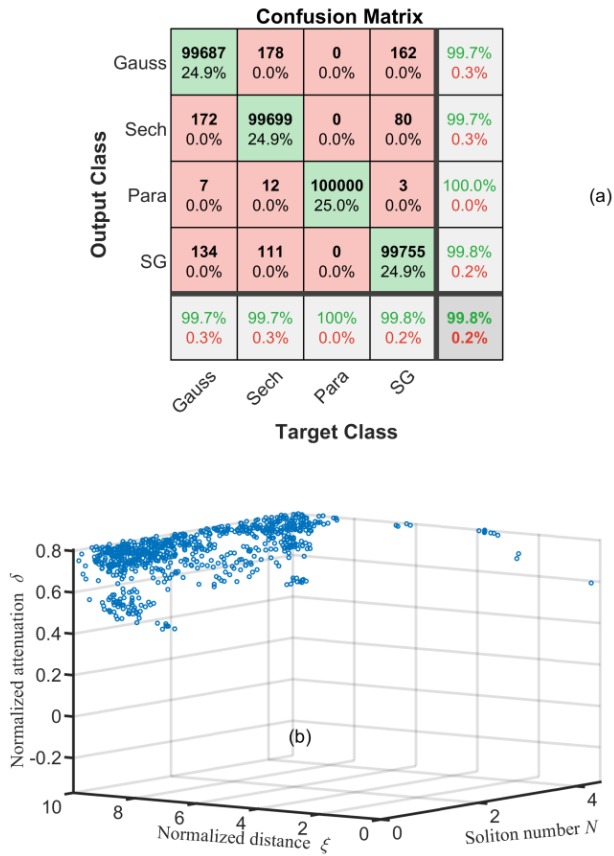

Figure 9: (a) Confusion matrix relating to the initial pulse shape when the NN is interrogated with randomly chosen new simulated output pulses from the fibre corresponding to an unlabeled mix

of input Gaussian, hyperbolic secant, parabolic and super-Gaussian pulses. (b) Points where prediction errors occur in the three-dimensional space ($N, \xi, \delta$).

## IV. Conclusion

In this paper, we have successfully applied machine learning to solve the direct and inverse problems relating to the shaping of optical pulses during nonlinear propagation in a normally dispersive fibre in the presence of distributed gain or loss, and our results expand previous studies of nonlinear pulse shaping using machine learning [24]. Using a feedforward NN trained on numerical simulations of the NLSE, we have shown that the temporal and spectral properties of the output pulses from the fibre can be predicted with high accuracy. The network is able to accommodate to and maintain high accuracy for a wide dynamic range of pulse parameters. The key properties of parabolic self-similar pulses are successfully reproduced by the NN model. Remarkably, within the range of system parameters considered, any temporal and spectral shape generated at the fibre output from the propagation of an initial Gaussian pulse can be associated with a single set of normalized propagation length, soliton-order number and gain/loss values, ($\xi$, $N, \delta$).

Our results show that a properly trained network can greatly help the characterization and inverse-engineering of fibre-based shaping systems by providing immediate and sufficiently accurate solutions. However, the presence of a global attractor in a fibre amplifier system for arbitrary initial pulse conditions can potentially introduce uncertainty in the inverse design problem or in the recognition of the input pulse characteristics. Although this remains to be studied in further detail, our results suggest that the use of NNs for pattern recognition in systems that possess asymptotic self-similar solutions may possess limitations. Note that we have also limited our discussion here to initially Fourier transform-limited pulses. Future steps may expand the parameter space of the NN operation for chirped initial pulses [14, 24]. The inclusion of higher-order propagation effects such as nonlinear longitudinal variation of the gain or gain saturation will add complexity to the nonlinear shaping problem by breaking the simple scaling laws from the NLSE. However, a NN algorithm should be able to tackle this problem and greatly accelerate its solution. Other future lines of research could be the analysis of the impact of noise on the accuracy of the NN predictions, and the deployment of convolutional NNs [13].

# Acknowledgements

C. F. and J.M.D were supported by the EUR EIPHI and I-SITE BFC projects (contracts ANR-17-EURE-0002 and ANR-15-IDEX-0003), the ANR OPTIMAL project, and the Région Bourgogne-Franche-Comté. C.F. was also supported by the Institut Universitaire de France. The numerical simulations relied on the HPC resources of DNUM CCUB (Centre de Calcul de l'Université de Bourgogne).